\documentclass[aps,prd,twocolumn,comment, showpacs,superscriptaddress,groupedaddress,floatfix,nofootinbib]{revtex4}  
\usepackage{float}
\usepackage{graphicx}
\usepackage{slashed}
\usepackage{hyperref}
\usepackage{amsmath}
\usepackage{dcolumn}   
\usepackage{bm}        
\usepackage{amssymb} 
\usepackage{color}

\begin{document}

%

\let\a=\alpha      \let\b=\beta       \let\c=\chi        \let\d=\delta
\let\e=\varepsilon \let\f=\varphi     \let\g=\gamma      \let\h=\eta
\let\k=\kappa      \let\l=\lambda     \let\m=\mu
\let\o=\omega      \let\r=\varrho     \let\s=\sigma
\let\t=\tau        \let\th=\vartheta  \let\y=\upsilon    \let\x=\xi
\let\z=\zeta       \let\io=\iota      \let\vp=\varpi     \let\ro=\rho
\let\ph=\phi       \let\ep=\epsilon   \let\te=\theta
\let\n=\nu
\let\D=\Delta   \let\F=\Phi    \let\G=\Gamma  \let\L=\Lambda
\let\O=\Omega   \let\P=\Pi     \let\Ps=\Psi   \let\Si=\Sigma
\let\Th=\Theta  \let\X=\Xi     \let\Y=\Upsilon
%


\def\cA{{\cal A}}                \def\cB{{\cal B}}
\def\cC{{\cal C}}                \def\cD{{\cal D}}
\def\cE{{\cal E}}                \def\cF{{\cal F}}
\def\cG{{\cal G}}                \def\cH{{\cal H}}
\def\cI{{\cal I}}                \def\cJ{{\cal J}}
\def\cK{{\cal K}}                \def\cL{{\cal L}}
\def\cM{{\cal M}}                \def\cN{{\cal N}}
\def\cO{{\cal O}}                \def\cP{{\cal P}}
\def\cQ{{\cal Q}}                \def\cR{{\cal R}}
\def\cS{{\cal S}}                \def\cT{{\cal T}}
\def\cU{{\cal U}}                \def\cV{{\cal V}}
\def\cW{{\cal W}}                \def\cX{{\cal X}}
\def\cY{{\cal Y}}                \def\cZ{{\cal Z}}


\def\be{\begin{equation}}
\def\ee{\end{equation}}
\def\bea{\begin{eqnarray}}
\def\eea{\end{eqnarray}}
\def\bm{\begin{matrix}}
\def\em{\end{matrix}}
\def\bpm{\begin{pmatrix}}
\def\epm{\end{pmatrix}}

\def\mpl{M_{\rm {Pl}}}
\def\tev{{\rm \,Te\kern-0.125em V}}
\def\gev{{\rm \,Ge\kern-0.125em V}}
\def\mev{{\rm \,Me\kern-0.125em V}}
\def\ev{\,{\rm eV}}

\title{\boldmath   Explaining the CMS excesses, baryogenesis and neutrino masses in $E_{6}$ motivated $U(1)_{N}$ model}
\author{Mansi Dhuria}
\email{mansi@prl.res.in}
\affiliation{Physical Research Laboratory, Navrangpura, Ahmedabad 380 009, India}
\author{Chandan Hati}
\email{chandan@prl.res.in} 
\affiliation{Physical Research Laboratory, Navrangpura, Ahmedabad 380 009, India}
\affiliation{Indian Institute of Technology Gandhinagar, Chandkheda, Ahmedabad 382 424, India.}
\author{Utpal Sarkar}
\email{utpal@prl.res.in} 
\affiliation{Physical Research Laboratory, Navrangpura, Ahmedabad 380 009, India}

\begin{abstract}
	We study the superstring inspired $E_{6}$ model motivated $U(1)_{N}$ extension of the supersymmetric standard model to explore the possibility of explaining  the recent excess CMS events and the baryon asymmetry of the universe in eight possible variants of the model. In light of the hints from short-baseline neutrino experiments at the existence of one or more light sterile neutrinos, we also study the neutrino mass matrices dictated by the field assignments and the discrete symmetries in these variants. We find that all the variants can explain the excess CMS events via the exotic slepton decay, while for a standard choice of the discrete symmetry four of the variants have the feature of allowing high scale baryogenesis (leptogenesis). For one other variant three body decay induced soft baryogenesis mechanism is possible which can induce baryon number violating neutron-antineutron oscillation. We also point out a new discrete symmetry which has the feature of ensuring proton stability and forbidding tree level flavor changing neutral current processes while allowing for the possibility of high scale leptogenesis for two of the variants. On the other hand, neutrino mass matrix of the $U(1)_{N}$ model variants naturally accommodates three active and two sterile neutrinos which acquire masses through their mixing with extra neutral fermions giving rise to interesting textures for neutrino masses.
\end{abstract}

\maketitle

\section{Introduction}
One of the simplest and well motivated extensions of the Standard Model (SM) gauge group $SU(3)_{c}\times SU(2)_{L} \times U(1)_{Y}$ is the $U(1)_{N}$ extension of the supersymmetric SM motivated by the superstring theory inspired $E_{6}$ model. This model, realizing the implementation of supersymmetry and the extension of the SM gauge group to a larger symmetry group, offers an attractive possibility of $\tev$-scale physics beyond the SM, testable at the LHC. On the other hand, small neutrino masses explaining the solar and atmospheric neutrino oscillations data and a mechanism for generating the observed baryon asymmetry of the universe can be naturally accommodated in this model. 

The presence of new exotic fields in addition to the SM fields and new interactions involving the new gauge boson $Z^{\prime}$ provides a framework to explore the associated rich phenomenology which can be tested at the LHC. To this end, we must mention that recently the CMS Collaboration at the LHC have reported excesses in the searches for the right-handed gauge boson $W_{R}$ at a center of mass energy of $\sqrt{s}=8 \tev$ and $19.7 \rm{fb}^{-1}$ of integrated luminosity \cite{Khachatryan:2014dka} and di-leptoquark production at a center of mass energy of $\sqrt{s}=8 \tev$ and $19.6 \rm{fb}^{-1}$ of integrated luminosity \cite{CMS:2014qpa}. In the former the final state $eejj$ was used to probe $pp\rightarrow W_{R}\rightarrow eN_{R}\rightarrow eejj$ and in the energy bin $1.8 \tev< m_{eejj}<2.2 \tev$ a $2.8\sigma$ local excess have been reported accounting for 14 observed events with 4 expected background events from the SM. In the search for di-leptoquark production, $2.4\sigma$ and $2.6\sigma$ local excesses in $eejj$ and $e\slashed{p}_{T}jj$ channels respectively have been reported corresponding to $36$ observed events with $20.49\pm 2.4\pm 2.45$(syst.) expected SM background events and $18$ observed events with $7.54\pm 1.20\pm 1.07$(syst.) expected SM background events respectively \cite{CMS:2014qpa}.

 Attempts have been made to explain the above CMS excesses in the context of Left-Right Symmetric Model (LRSM). The $eejj$ excess have been explained from $W_R$ decay for LRSM with $g_L = g_R$ by taking into account the CP phases and non-degenerate masses of heavy neutrinos in Ref. \cite{Gluza:2015goa}, and also by embedding the conventional LRSM with $g_L\ne g_R$ in the $SO(10)$ gauge group in Refs. \cite{Deppisch:2014qpa}. In these models, the lepton asymmetry can get generated either through the lepton number violating decay of right-handed Majorana neutrinos \cite{Fukugita:1986hr} or heavy Higgs triplet scalars \cite{Ma:1998dx}. However, the conventional LRSM models (even after embedding it in higher gauge groups) are not consistent with the canonical mechanism of leptogenesis in the range of the mass of $W_R$ $(\sim 2 \tev)$ corresponding to the $eejj$ excess at the LHC reported by the CMS \cite{Dhuria:2015cfa, Dhuria:2015wwa, Ma:1998sq}.
 
 The $eejj$ excess has also been discussed in the context of $W_R$ and $Z^\prime$ production and decay in Ref. \cite{Saavedra2014} and in the context of pair production of vector-like leptons in Refs. \cite{Dobrescu2014}. In Ref. \cite{strumia-2014}, a scenario connecting leptoquarks to dark matter was proposed accounting for the recent excess seen by CMS. In Refs. \cite{Allanach-2014, Biswas-2014}, the excess events have been shown to occur in $R$-parity violating processes via the resonant production of a slepton. In Ref. \cite{Dhuria:2015hta}, the three effective low-energy subgroups of the superstring inspired $E_{6}$ model with a low energy $SU(2)_{(R)}$ were studied and a $R$-parity conserving scenario was proposed in which both the $eejj$ and $e{\slashed p_T jj}$ signals can be produced from the decay of an exotic slepton in two of the effective low-energy subgroups of the superstring inspired $E_{6}$ model, out of which one subgroup (known as the Alternative Left-Right Symmetric Model \cite{Ma:1986we}) allows for the possibility of having successful high-scale leptogenesis.

 In this article, we systematically study the $E_{6}$ motivated $U(1)_{N}$ extension of the supersymmetric SM gauge group to explain the excess CMS events and simultaneously explain the baryon asymmetry of the universe via baryogenesis (leptogenesis). To this end, we impose discrete symmetries to the above gauge group which ensures proton stability, forbids the tree level flavor changing neutral current (FCNC) processes and dictates the form of the neutrino mass matrix in the variants of the $U(1)_{N}$ model. We find that all the variants can explain the excess CMS events via the exotic slepton decay, while for a standard choice of the discrete symmetry some of them have the feature of allowing high scale baryogenesis (leptogenesis) via the decay of a heavy Majorana baryon (lepton) and some are not consistent with such mechanisms. We have pointed out the possibility of the three body decay induced soft baryogenesis mechanism which can induce baryon number violating neutron-antineutron ($n-\bar{n}$) oscillation \cite{Mohapatra:1980qe} in one such variant, on the other hand, we have also explored a new discrete symmetry for these variants which has the feature of ensuring proton stability and forbidding tree level FCNC processes while allowing for the possibilities of high scale leptogenesis through the decay of a heavy Majorana lepton. In light of the hints from short-baseline neutrino experiments \cite{Aguilar:2001ty} at the existence of one  or more light sterile neutrinos which can interact only via mixing with the active neutrinos, we have also explored the neutrino mass matrix of the $U(1)_{N}$ model variants which naturally contains three active and two sterile neutrinos \cite{Ma:1995xk}. These neutrinos acquire masses through their mixing with extra neutral fermions giving rise to interesting textures for neutrino masses governed by the field assignments and the imposed discrete symmetries.

The outline of the article is as follows.  In Sec. {\ref{sec_E6subgroups}}, we review the $E_{6}$ model motivated $U(1)_{N}$ extension of supersymmetric standard model and the transformations of the various superfields under the gauge group. In Sec. {\ref{u1N_models}}, we discuss the imposition of discrete symmetries and give the variants of the $U(1)_{N}$ model and the corresponding superpotentials. In Sec. {\ref{eejj}} we discuss the possibility of producing $eejj$ and $e{\slashed p_T jj}$ events from the decay of an exotic slepton.  In Sec. {\ref{leptogenesis}}, we explore the possible mechanisms of baryogenesis (leptogenesis) for the different variants of the $U(1)_{N}$ model. In Sec. {\ref{nmass}}, we study the neutral fermionic mass matrices and the resultant structure of the neutrino mass matrices. In Sec. {\ref{conclusions}} we conclude with our results.

\section{$U(1)_{N}$ extension of supersymmetric Standard Model}
\label{sec_E6subgroups}
In the heterotic superstring theory with $E_8 \times E'_8$ gauge group the compactification on a Calabi-Yau manifold leads to the breaking of $E_8$ to $SU(3) \times E_6$ \cite{Candelasetal1985, Greene:1996cy}. The flux breaking of $E_6$ can result in different low-energy effective subgroups of rank-5 and rank-6. One such possibility is realized in the $U(1)_{N}$ model. The rank - 6 group $E_{6}$ can be broken down to low-energy gauge groups of rank - 5 or rank - 6 with one or two additional $U(1)$ in addition to the SM gauge group. For example $E_{6}$ contains the subgroup $SO(10)\times U(1)_{\psi}$ while $SO(10)$ contains the subgroup $SU(5)\times U(1)_{\chi}$. In fact some mechanisms can break the $E_{6}$ group directly into the rank -6 gauge scheme
\be{\label{1.0.1}}
E_{6} \rightarrow SU(3)_{C}\times SU(2)_{L}\times U(1)_{Y}\times U(1)_{\psi} \times U(1)_{\chi}.
\ee
These rank - 6 schemes can further be reduced to rank - 5 gauge group with only one additional $U(1)$ which is a linear combination of $U(1)_{\psi}$ and $U(1)_{\chi}$
\be{\label{1.0.2}}
Q_{\alpha}=Q_{\psi} \cos \alpha +Q_{\chi} \sin \alpha,
\ee
where
\be{\label{1.0.3}}
Q_{\psi}=\sqrt{\frac{3}{2}} (Y_{L}-Y_{R}), \; \; Q_{\chi}=\sqrt{\frac{1}{10}} (5T_{3R}-3Y).
\ee
For a particular choice of $\tan \alpha=\sqrt{\frac{1}{15}}$  the right-handed counter part of neutrino superfield ($N^{c}$) can transform trivially under the gauge group and the corresponding $U(1)$ gauge extension to the SM is denoted as $U(1)_{N}$. The trivial transformation of $N^{c}$ can allow a large Majorana mass of $N^{c}$ in the $U(1)_{N}$ model thus providing attractive possibility of baryogenesis (leptogenesis).

Let us consider one of the maximal subgroups of $E_{6}$ given by $SU(3)_{C}\times SU(3)_{L} \times SU(3)_{R}$. The fundamental $27$ representation of $E_{6}$ under this subgroup is given by
\be{\label{1.1}} 
27= (3, 3, 1)+(3^{\ast}, 1,3^{\ast})+(1, 3^{\ast},3)
\ee
The matter superfields of the first family are assigned as:
\be{\label{1.2}}
 \bpm u \\ d \\ h \epm + \bpm u^{c} & d^{c} &h^{c}\epm +\bpm E^{c} & \nu & \nu_{E} \\ N^{c}_{E} & e & E \\ e^{c} & N^{c} & n \epm,
\ee
where $SU(3)_{L}$ operates vertically and $SU(3)_{R}$ operates horizontally. Now if the $SU(3)_{L}$ gets broken to $SU(2)_{L}\times U(1)_{Y_{L}}$ and the $SU(3)_{R}$ gets broken to $U(1)_{T_{3R}}\times U(1)_{Y_{R}}$ via the flux mechanism then the resulting gauge symmetry is given by $SU(3)_{C}\times SU(2)_{L}\times U(1)_{Y}\times U(1)_{N}$, where the $U(1)_{N}$ charge assignment is given by
\be{\label{1.3}}
Q_{N}=\sqrt{\frac{1}{40}}(6Y_{L}+T_{3R}-9Y_{R}), 
\ee
and the electric charge is given by
\be{\label{1.4}}
Q=T_{3L}+Y, \; \; Y=Y_{L}+T_{3R}+Y_{R}.
\ee
The transformations of the various superfields of the fundamental 27 representation of $E_{6}$ under $SU(3)_{C}\times SU(2)_{L}\times U(1)_{Y}\times U(1)_{N}$ and the corresponding assignments of $Y_{L}$, $T_{3R}$ and $Y_{R}$ are listed in Table \ref{tab:table1}, where $Q=(u, d)$, $L=(\nu_{e}, e)$, $X=(\nu_{E}, E)$ and $X^{c}=(E^{c}, N_{E}^{c})$.
\begin{table}
	\caption{\label{tab:table1} Transformations of the various superfields of the 27 representation under $SU(3)_{C}\times SU(2)_{L}\times U(1)_{Y}\times U(1)_{N}$.}
	\begin{ruledtabular}
	\begin{tabular}{cccccccc}
			& $SU(3)_{c}$ & $SU(2)_{L}$ & $Y_{L}$ & $T_{3R}$ & $Y_{R}$ & $U(1)_{Y}$ & $U(1)_{N}$\\
			\hline
			\\
			\vspace {0.05in}
			$Q$ & 3 & 2 & $\frac{1}{6}$ & 0 & 0 & $\frac{1}{6}$ & $\frac{1}{\sqrt{40}}$\\
			 \vspace {0.05in}
			$u^{c}$ & $3^{\ast}$ & 1 & 0 & $-\frac{1}{2}$ & $-\frac{1}{6}$ &$ -\frac{2}{3}$ & $\frac{1}{\sqrt{40}}$\\
			 \vspace {0.05in}
			$d^{c}$ & $3^{\ast}$ & 1 & 0 & $\frac{1}{2}$ & $-\frac{1}{6}$ & $\frac{1}{3}$ & $\frac{2}{\sqrt{40}}$\\
			\vspace {0.05in}
			        $L$ & 1 & 2 & $-\frac{1}{6}$ & 0 & $-\frac{1}{3}$ & $-\frac{1}{2}$ & $\frac{2}{\sqrt{40}}$\\
			        \vspace {0.05in}
			$e^{c}$ & 1 & 1 & $\frac{1}{3}$ & $\frac{1}{2}$ & $\frac{1}{6}$ & 1 & $\frac{1}{\sqrt{40}}$\\
			\vspace {0.05in}
			        $h$ & 3 & 1 & $-\frac{1}{3}$ & 0 & 0 & $-\frac{1}{3}$ & $-\frac{2}{\sqrt{40}}$\\
			        \vspace {0.05in}
			$h^{c}$ & $3^{\ast}$ & 1 & 0 & 0 & $\frac{1}{3}$ & $\frac{1}{3}$ & $-\frac{3}{\sqrt{40}}$\\
			\vspace {0.05in}
			      $X$ & 1 & 2 & $-\frac{1}{6}$ & $-\frac{1}{2}$ & $\frac{1}{6}$ & $-\frac{1}{2}$ & $-\frac{3}{\sqrt{40}}$\\
			      \vspace {0.05in}
			$X^{c}$ & 1 & 2 &$ -\frac{1}{6}$ & $\frac{1}{2}$ & $\frac{1}{6}$ & $\frac{1}{2}$ & $-\frac{2}{\sqrt{40}}$\\
			\vspace {0.05in}
			        $n$ & 1 & 1 & $\frac{1}{3}$ & 0 & $-\frac{1}{3}$ & 0 & $\frac{5}{\sqrt{40}}$\\
			        \vspace {0.05in}
			$N^{c}$ & 1 & 1 & $\frac{1}{3}$ & $-\frac{1}{2}$ & $\frac{1}{6}$ & 0 & 0\\
			\end{tabular}
	\end{ruledtabular}
 \end{table}
\section{Discrete symmetries and variants of $U(1)_{N}$ model}
\label{u1N_models}
The presence of the extra particles in this model can have interesting phenomenological consequences, however, they can also cause serious problems regarding fast proton decay, tree level flavor changing neutral current (FCNC) and neutrino masses. Considering the decomposition of $27\times 27 \times 27$ there are eleven possible superpotential terms. The most general superpotential can be written as
\bea{\label{2.1}}
W &=& W_{0}+W_{1}+W_{2}, \nonumber \\
W_{0}&=& \l_{1} Qu^{c} X^{c}+ \l_{2} Qd^{c}X+ \l_{3} Le^{c}X + \nonumber \\
            && \l_{4} Shh^{c}+\l_{5} SXX^{c}+ \l_{6} LN^{c}X^{c}+\l_{7} d^{c}N^{c}h,\nonumber \\
W_{1}&=& \l_{8} QQh +\l_{9} u^{c}d^{c}h^{c},\nonumber \\
W_{2}&=& \l_{10} QLh^{c} +\l_{11} u^{c}e^{c}h .
\eea
The first five terms of $W_{0}$ give masses to the usual SM particles and the new heavy particles $h, h^{c}, X$ and $X^{c}$. The last term of $W_{0}$ i.e. $LN^{c}X^{c}$ can generate a non zero Dirac neutrino mass and in some scenarios it is desirable to have the coupling $\l_{6}$ very small or vanishing, so that the three neutrinos pick up small masses. Now the rest five terms corresponding to $W_{1}$ and $W_{2}$ cannot all be there together as it would induce rapid proton decay. Imposition of a discrete symmetry can forbid such terms and give a sufficiently longlived proton \cite{Joshipura:1986mt}. We will impose a $Z^{B}_{2}\times Z^{H}_{2}$ discrete symmetry, where the first $Z^{B}_{2}=(-1)^{3B}$ prevents rapid proton decay and the second discrete symmetry $Z^{H}$ distinguishes between the Higgs and matter supermultiplets and suppress the tree level FCNC processes.
\begin{table}
	\caption{\label{tab:table2} Possible transformations of $h,h^{c}$ and $N^{c}$ under $Z^{B}_{2}$ and the allowed superpotential terms.}
	\begin{ruledtabular}
		\begin{tabular}{cccc}
			Model & $h,h^{c}$ & $N^{c}$ & Allowed trilinear terms \\
			\hline
			\\
			1 & +1 &-1& $W_{0}$ ($\l_{6}=0$), $W_{1}$\nonumber\\
			\\
			2 & +1 &-1 for $N^{c}_{1,2}$,& $W_{0}$ ($\l_{6}=0$ for $N^{c}_{1,2}$, \nonumber\\
			&&+1 for $N^{c}_{3}$&$\l_{7}=0$ for $N^{c}_{3}$),$W_{1}$\nonumber\\
			\\
			3 & -1 &+1& $W_{0}$, $W_{2}$\nonumber\\
			\\
			4 & -1 &+1 for $N^{c}_{1,2}$,& $W_{0}$ ($\l_{6}=\l_{7}=0$ for $N^{c}_{3}$), $W_{2}$\nonumber\\
			&&-1 for $N^{c}_{3}$&\nonumber\\
			\\
			5 & +1 &+1 for $N^{c}_{1,2}$,& $W_{0}$ ($\l_{6}=0$ for $N^{c}_{3}$,\nonumber\\
			&&-1 for $N^{c}_{3}$&$\l_{7}=0$ for $N^{c}_{1,2}$), $W_{1}$\nonumber\\
			\\
			6 & +1 &+1& $W_{0}$ ($\l_{7}=0$), $W_{1}$\nonumber\\
			\\
			7 & -1 &-1& $W_{0}$ ($\l_{6}=\l_{7}=0$), $W_{2}$\nonumber\\
			\\
			8 & -1 &-1 for $N^{c}_{1,2}$,& $W_{0}$ ($\l_{6}=\l_{7}=0$ for $N^{c}_{1,2}$), $W_{2}$\nonumber\\
			&&+1 for $N^{c}_{3}$&\nonumber\\
		\end{tabular}
	\end{ruledtabular}
\end{table}

Under $Z^{B}_{2}=(-1)^{3B}$ we have
\bea{\label{2.2}}
Q, u^{c}, d^{c} \; &:& \; -1 \nonumber\\
L, e^{c}, X, X^{c}, S \; &:& \; +1,
\eea
now depending on the assignments of $h, h^{c}$ and $N^{c}$ one can have different variants of the model. Such different possibilities are listed in Table \ref{tab:table2}.

In the models where $h, h^{c}$ are even under $Z^{B}_{2}$ the superfields $h (B=-2/3)$ and $h^{c} (B=2/3)$ are diquarks while for the rest $h (B=1/3, L=1)$ and $h^{c} (B=-1/3, L=-1)$ are leptoquarks. $N^{c}$ with the assignment $Z^{B}_{2}=-1$ are baryons and the assignment $Z^{B}_{2}=+1$ are leptons. In addition to the trilinear terms listed in Table \ref{tab:table2} there can be bilinear terms such as $LX^{c}$ and $N^{c}N^{c}$. The former can give rise to nonzero neutrino mass and the latter can give heavy Majorana baryon (lepton) $N^{c}$ mass. Model 1 is similar to model 5 of Ref. \cite{Ma:1987jg} and model A of Ref. \cite{Ma:2000jf}. Model 2 is same as model B of Ref. \cite{Ma:2000jf}. Model 8 is quite different from the ones that have been discussed in connection with leptogenesis in the literature (e.g. \cite{ma_2000}). Here the matter superfields $X,X^{c}$ carry non zero $B-L$ quantum numbers and the tree level FCNC processes are forbidden. 

\subsection{Model 1}
In this model we take the second discrete symmetry $Z^{H}_{2}$  to be $Z^{L}_{2}=(-1)^{L}$ following Ref. \cite{Ma:2000jf} and it is imposed as follows
\bea{\label{2.3}}
L, e^{c}, X_{1,2}, X^{c}_{1,2}, S_{1,2}\; &:& \; -1 \nonumber\\
Q, u^{c}, d^{c}, N^{c}, h, h^{c}, S_{3}, X_{3}, X^{c}_{3}  \; &:& \; +1.
\eea
The neutral Higgs superfields $S_{3}, X_{3}$ and $X^{c}_{3}$ have zero lepton numbers and can pick up vacuum expectation values (VEVs) while the presence of the bilinear terms $LX^{c}_{1,2}$ imply that $X^{c}_{1,2}$ have $L=-1$ and $X_{1,2}$ have $L=1$. In this model $N^{c}$ is a baryon with $B=1$ and it acquires a Majorana mass from the bilinear term $mN^{c}N^{c}$. The complete superpotential of model 1 is given by
\bea{\label{2.4}}
W&=& \l_{1}^{ij} Q_{j}u^{c}_{i} X^{c}_{3}+ \l_{2}^{ij} Q_{j}d^{c}_{i}X_{3}+ \l_{3} L_{j}e^{c}_{i}X_{3} +  \l_{4}^{ij} S_{3}h_{i}h^{c}_{j}\nonumber \\
&+&\l_{5}^{3ab} S_{3}X_{a}X^{c}_{b}+\l_{5}^{a3b} S_{a}X_{3}X^{c}_{b}+\l_{5}^{ab3} S_{a}X_{b}X^{c}_{3}\nonumber \\
&+&\l_{5}^{333} S_{3}X_{3}X^{c}_{3} +\l_{7}^{ijk} d^{c}_{i}h_{j}N^{c}_{k}+ \m^{ia}L_{i}X^{c}_{a}\nonumber \\
&+& m^{ij}_{N} N^{c}_{i} N^{c}_{j} + W_{1},
\eea
where $i, j, k$ are flavor indices which run over all 3 flavors and $a, b=1, 2$ \footnote{We will use this notation hereafter in this article. The indices $i, j, k$ run over 1,2,3,  while the indices $a, b$ run over 1,2.}. The form of the superpotential clearly shows that the up-type quarks couple to $X^{c}_{3}$ only while the down-type quarks and the charged leptons couple to $X_{3}$ only, resulting in the suppression of the FCNC processes at the tree level. 

\subsection{Model 2}
Here the second discrete symmetry $Z^{L}_{2}$ is imposed as follows
\bea{\label{2.2.1}}
L, e^{c}, X_{1,2}, X^{c}_{1,2}, S_{1,2}, N^{c}_{3}\; &:& \; -1 \nonumber\\
Q, u^{c}, d^{c}, N^{c}_{1,2}, h, h^{c}, S_{3}, X_{3}, X^{c}_{3}  \; &:& \; +1.
\eea
In this model $N^{c}_{1,2}$ are baryons with $B=1$ but $N^{c}_{3}$ is a lepton and can give mass to one of the neutrinos via the term $LN^{c}_{3}X^{c}_{3}$. The complete superpotential of model 2 is given by
\bea{\label{2.2.2}}
W&=& \l_{1}^{ij} Q_{j}u^{c}_{i} X^{c}_{3}+ \l_{2}^{ij} Q_{j}d^{c}_{i}X_{3}+ \l_{3} L_{j}e^{c}_{i}X_{3} +  \l_{4}^{ij} S_{3}h_{i}h^{c}_{j}\nonumber \\
&+&\l_{5}^{3ab} S_{3}X_{a}X^{c}_{b}+\l_{5}^{a3b} S_{a}X_{3}X^{c}_{b}+\l_{5}^{ab3} S_{a}X_{b}X^{c}_{3}\nonumber\\
&+&\l_{5}^{333} S_{3}X_{3}X^{c}_{3} + \l_{6}^{i} L_{i}N^{c}_{3}X^{c}_{3} +\l_{7}^{ija} d^{c}_{i}h_{j}N^{c}_{a} +  \m^{ia}L_{i}X^{c}_{a}\nonumber\\
&+& m^{ab}_{N} N^{c}_{a} N^{c}_{b}+ m^{33}_{N} N^{c}_{3} N^{c}_{3} + W_{1}.
\eea

\subsection{Model 3}
Under the second discrete symmetry $Z^{H}_{2}=Z^{L}_{2}=(-1)^{L}$ the superfields transform as follows
\bea{\label{2.4.1}}
L, e^{c}, X_{1,2}, X^{c}_{1,2}, S_{1,2}, N^{c}, h, h^{c}\; &:& \; -1 \nonumber\\
Q, u^{c}, d^{c}, S_{3}, X_{3}, X^{c}_{3}  \; &:& \; +1.
\eea
In this model all the $N^{c}$s are leptons. The complete superpotential of model 4 is given by
\bea{\label{2.4.2}}
W&=& \l_{1}^{ij} Q_{j}u^{c}_{i} X^{c}_{3}+ \l_{2}^{ij} Q_{j}d^{c}_{i}X_{3}+ \l_{3} L_{j}e^{c}_{i}X_{3} +  \l_{4}^{ij} S_{3}h_{i}h^{c}_{j}\nonumber \\
&+&\l_{5}^{3ab} S_{3}X_{a}X^{c}_{b}+\l_{5}^{a3b} S_{a}X_{3}X^{c}_{b}+\l_{5}^{ab3} S_{a}X_{b}X^{c}_{3}\nonumber\\
&+&\l_{5}^{333} S_{3}X_{3}X^{c}_{3} + \l_{6}^{ij3} L_{i}N^{c}_{j}X^{c}_{3} +\l_{7}^{ijk} d^{c}_{i}h_{j}N^{c}_{k}\nonumber\\
&+& \m^{ia}L_{i}X^{c}_{a}+ m^{ij}_{N} N^{c}_{i} N^{c}_{j} + W_{2}.
\eea

\subsection{Model 4}
Here the second discrete symmetry $Z^{H}_{2}$ is again chosen to be $(-1)^{L}$ giving the transformations of the superfields as follows
\bea{\label{2.3.1}}
L, e^{c}, X_{1,2}, X^{c}_{1,2}, S_{1,2}, N^{c}_{1,2}, h, h^{c}\; &:& \; -1 \nonumber\\
Q, u^{c}, d^{c}, N^{c}_{3}, S_{3}, X_{3}, X^{c}_{3}  \; &:& \; +1.
\eea
$N^{c}_{1,2}$ are leptons while $N^{c}_{3}$ is a baryon. The complete superpotential of model 2 is given by
\bea{\label{2.3.2}}
W&=& \l_{1}^{ij} Q_{j}u^{c}_{i} X^{c}_{3}+ \l_{2}^{ij} Q_{j}d^{c}_{i}X_{3}+ \l_{3} L_{j}e^{c}_{i}X_{3} +  \l_{4}^{ij} S_{3}h_{i}h^{c}_{j}\nonumber \\
&+&\l_{5}^{3ab} S_{3}X_{a}X^{c}_{b}+\l_{5}^{a3b} S_{a}X_{3}X^{c}_{b}+\l_{5}^{ab3} S_{a}X_{b}X^{c}_{3}\nonumber\\
&+&\l_{5}^{333} S_{3}X_{3}X^{c}_{3} + \l_{6}^{ia3} L_{i}N^{c}_{a}X^{c}_{3} +\l_{7}^{ija} d^{c}_{i}h_{j}N^{c}_{a}\nonumber\\
&+& \m^{ia}L_{i}X^{c}_{a}+ m^{ab}_{N} N^{c}_{a} N^{c}_{b}+ m^{33}_{N} N^{c}_{3} N^{c}_{3} +W_{2}.
\eea

\subsection{Model 5 and 6}
In model 5 if we choose the second discrete symmetry $Z^{H}_{2}$ to be $Z^{L}_{2}=(-1)^{L}$ then the superfields transform as follows
\bea{\label{2.5.1}}
L, e^{c}, X_{1,2}, X^{c}_{1,2}, S_{1,2}, N^{c}_{1,2}\; &:& \; -1 \nonumber\\
Q, u^{c}, d^{c}, N^{c}_{3}, h, h^{c}, S_{3}, X_{3}, X^{c}_{3}  \; &:& \; +1,
\eea
which forbids the terms $\l_{6} L_{i}N^{c}_{a}X^{c}_{b}$ ($\l_{7}$ is already vanishing for $N^{c}_{1,2}$ from the imposition of the first discrete symmetry $Z^{B}_{2}$) and thus the possibility of high scale baryogenesis (via leptogenesis) through the decay of Majorana $N^{c}$ gets ruled out. However there can be soft baryogenesis through three body decays which can induce $n-\bar{n}$ oscillation. We will elaborate on this in Section \ref{leptogenesis}. With the above choice of second discrete symmetry given in Eq. (\ref{2.5.1}) the complete superpotential for model 5 is given by
\bea{\label{2.5.2}}
W&=& \l_{1}^{ij} Q_{j}u^{c}_{i} X^{c}_{3}+ \l_{2}^{ij} Q_{j}d^{c}_{i}X_{3}+ \l_{3} L_{j}e^{c}_{i}X_{3} +  \l_{4}^{ij} S_{3}h_{i}h^{c}_{j}\nonumber \\
&+&\l_{5}^{3ab} S_{3}X_{a}X^{c}_{b}+\l_{5}^{a3b} S_{a}X_{3}X^{c}_{b}+\l_{5}^{ab3} S_{a}X_{b}X^{c}_{3}\nonumber\\
&+&\l_{5}^{333} S_{3}X_{3}X^{c}_{3} + \l_{6}^{ia} L_{i}N^{c}_{a}X^{c}_{3} +\l_{7}^{ij3} d^{c}_{i}h_{j}N^{c}_{3} +  \m^{ia}L_{i}X^{c}_{a}\nonumber\\
&+& m^{ab}_{N} N^{c}_{a} N^{c}_{b}+ m^{33}_{N} N^{c}_{3} N^{c}_{3} + W_{1}.
\eea

We find that in this model it is possible to allow high scale leptogenesis through the decay of Majorana $N^{c}$ by a clever choice of the second discrete symmetry such that it can distinguish between the matter and Higgs superfields and also suppress the unwanted FCNC processes at the tree level. One such choice can be $Z^{E}_{2}$ which is associated with most of the exotic states. We define the transformation properties of the various superfields under $Z^{H}_{2}=Z^{E}_{2}$ as follows
\bea{\label{2.5.3}}
X_{1,2}, X^{c}_{1,2}, S_{1,2}, N^{c} \; &:& \; -1 \nonumber\\
L, e^{c}, Q, u^{c}, d^{c}, h, h^{c}, S_{3}, X_{3}, X^{c}_{3}  \; &:& \; +1,
\eea
Thus for this choice also $X_{3}, X^{c}_{3}$ and $S_{3}$ are the Higgs superfields that acquire VEVs. Since up-type quarks couple to $X^{c}_{3}$ only and down-type quarks and charged SM leptons couple to only $X_{3}$ the FCNC processes at the tree level are suppressed. The complete superpotential of model 5 with the assignments in Eq. \ref{2.5.3} reduces to
\bea{\label{2.5.4}}
W^{\prime}&=& \l_{1}^{ij} Q_{j}u^{c}_{i} X^{c}_{3}+ \l_{2}^{ij} Q_{j}d^{c}_{i}X_{3}+ \l_{3} L_{j}e^{c}_{i}X_{3} +  \l_{4}^{ij} S_{3}h_{i}h^{c}_{j}\nonumber \\
&+&\l_{5}^{3ab} S_{3}X_{a}X^{c}_{b}+\l_{5}^{a3b} S_{a}X_{3}X^{c}_{b}+\l_{5}^{ab3} S_{a}X_{b}X^{c}_{3}\nonumber\\
&+&\l_{5}^{333} S_{3}X_{3}X^{c}_{3} + \l_{6}^{iab} L_{i}N^{c}_{a}X^{c}_{b}\nonumber\\
&+& m^{ab}_{N} N^{c}_{a} N^{c}_{b} + m^{33}_{N} N^{c}_{3} N^{c}_{3} + W_{1}.
\eea

In model 6 also, the similar assignments for the superfields as given in Eq. (\ref{2.5.3}) holds good and the complete superpotential is similar to Eq. (\ref{2.5.4}) except the $\l_{6}$ term which now reads $\l_{6}^{ija} L_{i}N^{c}_{j}X^{c}_{a}$.

\subsection{Model 7 and 8}
Taking second discrete symmetry to be $Z^{H}_{2}=(-1)^{L}$ the superfields transform as follows
\bea{\label{2.7.1}}
L, e^{c}, X_{1,2}, X^{c}_{1,2}, S_{1,2}, h, h^{c}\; &:& \; -1 \nonumber\\
Q, u^{c}, d^{c}, N^{c}, S_{3}, X_{3}, X^{c}_{3}  \; &:& \; +1.
\eea
In this model all the $N^{c}$s are baryons. The complete superpotential of model 7 is given by
\bea{\label{2.7.2}}
W&=& \l_{1}^{ij} Q_{j}u^{c}_{i} X^{c}_{3}+ \l_{2}^{ij} Q_{j}d^{c}_{i}X_{3}+ \l_{3} L_{j}e^{c}_{i}X_{3} +  \l_{4}^{ij} S_{3}h_{i}h^{c}_{j}\nonumber \\
&+&\l_{5}^{3ab} S_{3}X_{a}X^{c}_{b}+\l_{5}^{a3b} S_{a}X_{3}X^{c}_{b}+\l_{5}^{ab3} S_{a}X_{b}X^{c}_{3}\nonumber\\
&+&\l_{5}^{333} S_{3}X_{3}X^{c}_{3}+\m^{ia}L_{i}X^{c}_{a}+ m^{ij}_{N} N^{c}_{i} N^{c}_{j} + W_{2}.
\eea
 Note that the $\l_{6}$ and $\l_{7}$ terms which are essential for baryogenesis through $N^{c}$ decay (as discussed in Section \ref{leptogenesis}) are forbidden by the $Z_{2}^{B}$ symmetry irrespective of what $Z_{2}^{H}$ one chooses. For model 8 also one can write down the superfield transformations and the superpotential. In this case the mass term for $N^{c}$ is given by $m^{ab}_{N} N^{c}_{a} N^{c}_{b}+ m^{33}_{N} N^{c}_{3} N^{c}_{3}$ and the terms $\l_{6}^{i33} L_{i}N^{c}_{3}X^{c}_{3}$, $\l_{7}^{ij3} d^{c}_{i}h_{j}N^{c}_{3}$ are present in addition to the terms given in Eq. (\ref{2.7.2}).
\section{Explaining the CMS $eejj$ (and $e\slashed{p}_{T}jj$) excess(es)}
\label{eejj}
\begin{figure}[h]
	\includegraphics[width=9cm]{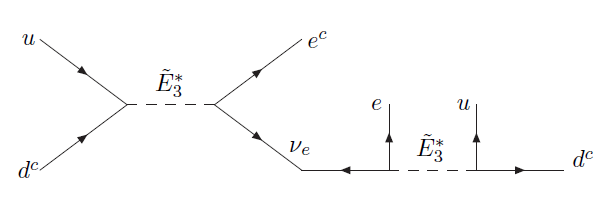}
	\caption{Feynman diagram for a  single exotic particle ${\tilde E}$ production leading to $eejj$ signal.}
	\label{fig:eejj2}
\end{figure}

An inspection of Table \ref{tab:table2} and the corresponding allowed superpotential terms reveals that all the models listed there contain the terms $\l_{2} Q_{i}d^{c}_{j}X_{3}$ and $\l_{3} L_{i}e^{c}_{j}X_{3}$ in the superpotential ($\tilde{N}^{c}_{E}$ and $\tilde{\n}_{E}$ acquires VEVs and $SU(2) \times U(1)_{Y}$ gets broken to $U(1)_{\rm{EM}}$) and can give rise to $eejj$ signal from the exotic slepton ${\tilde E}$ decay.  ${\tilde E}$ can be resonantly produced in $pp$ collisions, which then subsequently decays to a charged lepton and neutrino, followed by interactions of the neutrino producing an $eejj$ signal. The process leading to  $eejj$ signal is given in Fig. \ref{fig:eejj2}.

\begin{figure}[h]
	\includegraphics[width=9cm]{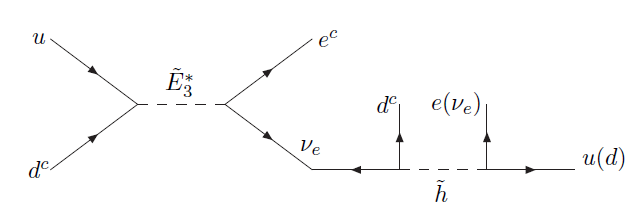}
	\caption{Feynman diagram for exotic slepton ${\tilde E}$ production leading to both $eejj$ and $e {\slashed p_T}jj$ signal}
	\label{fig:eejj1}
\end{figure}

The models where $h$ and $h^{c}$ are leptoquarks (Models 3, 4, 7 and 8 in Table \ref{tab:table2}) can produce both $eejj$ and $e\slashed p_T jj$ signals from the decay of scalar superpartner(s) of the exotic particle(s). Both events can be produced in the above scenarios via (i) resonant production of the exotic slepton $\tilde E$ (ii) and pair production of scalar leptoquarks ${\tilde h}$. The processes involving exotic slepton decay leading to both $eejj$ and $e\slashed p_T jj$ signals are given in Fig. \ref{fig:eejj1}. The superpotential terms involved in these processes are $\l_{10} QLh^{c}$ and $\l_{11} u^{c}e^{c}h$ in addition the two terms responsible for the first signal.
The  partonic cross section of slepton production is given by \cite{Allanach-2009}
\be
{\hat \sigma}=\frac{\pi}{ 12 \hat s} \left| \l_{2}\right|^2 \delta (1-\frac{ m^{2}_{\tilde E}}{\hat s}),
\ee
where ${\hat s}$ is the partonic center of mass energy, and $m_{\tilde E}$ is the mass of the resonant slepton. The total cross section is approximated to be \cite{Allanach-2009}
\be
\sigma\left(pp \rightarrow eejj\right) \propto \frac{\left | \lambda_{2}\right|^2}{m^{3}_{\tilde E}}  \times \beta_1  \ee
and
\be
\sigma\left(pp \rightarrow e {\slashed p_T}j j\right) \propto \frac{\left | \lambda_{2}\right|^2}{m^{3}_{\tilde E}}  \times  \beta_2, 
\ee
where $\beta_1$ is the branching fraction for the decay of $\tilde {E}$ to $eejj$ and $\beta_2$ is the branching fraction for the decay to $e\slashed p_T jj$. $\beta_{1,2}$ and the coupling $\lambda_{2}$ are the free parameters. The cross section for the processes can be calculated as a function of the exotic slepton mass and  bounds for the value of the mass of the exotic slepton  can be obtained by matching the theoretically calculated excess events with the ones observed at the LHC at a center of mass energy $\sqrt{s}=8 \tev$. Thus, the $U(1)_{N}$ models can explain the excess   $eejj$ (and $e {\slashed p_T}jj$) signal(s) at the LHC via resonant  exotic slepton decay.
\section{Baryogenesis (leptogenesis) in $U(1)_{N}$ models}
\label{leptogenesis}
Some of the variants of low-energy $U(1)_{N}$ subgroup of $E_6$ model allows for the possibility of explaining baryogenesis (leptogenesis)  from the decay of heavy Majorana particle $N^{c}$. In order to generate the baryon asymmetry of the Universe from $N^{c}$ decay the conditions that must be satisfied are (i) violation of $B-L$ from Majorana mass of $N^{c}$, (ii)  complex couplings must give rise to sufficient $CP$ violation and (iii) the out-of-equilibrium condition given by
\be{\label{4.1.1}}
\Gamma_{N}< H(T=m_{N})=\sqrt{\frac{4\pi^{3}g_{\ast}}{45}} \frac{T^{2}}{M_{Pl}},
\ee
must be satisfied, where $\Gamma_{N}$ is the  decay width of Majorana $N^{c}$, $H(T)$ is the Hubble rate, $g_{\ast}$ is the effective number of relativistic degrees of freedom at temperature $T$ and $M_{Pl}$ is the Planck mass. This implies that $N^{c}$ cannot transform nontrivially under the low-energy subgroup $G= SU(3)_{C}\times SU(2)_{L}\times U(1)_{Y}\times U(1)_{N}$, which is readily satisfied in some variants of $U(1)_{N}$ model (see Table \ref{tab:table1}).  Thus the out-of-equilibrium decay of heavy $N^{c}$ can give rise to high-scale baryogenesis (leptogenesis).

\begin{figure}[h]
	\includegraphics[width=9cm]{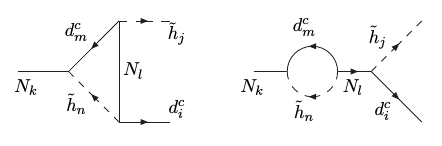}
	\caption{One-loop diagrams for $N_k$ decay which interferes with the tree level decay to provide $CP$ violation.}
	\label{fig:baryogenesis1}
\end{figure}

Models 1 and 2 have  distinctive features of allowing direct baryogenesis via decay of heavy Majorana baryon $N^{c}$ \cite{Ma:2000jf}. In both schemes, $N^c_{k (a)}$ decays to $B-L=B=-1$ final states $d^{c}_{i} {\tilde h}_{j}, {\tilde d^{c}}_{i} h_{j}$ and to their conjugate states with $B-L=B=1$, via the interaction term $\lambda^{ijk}_{7}$ ($\lambda^{ija}_{7}$ ) in Eq. (\ref{2.4} (\ref{2.2.2})). In both cases, the $CP$ violation comes from the complex Yukawa  coupling $\lambda^{ijk}_{7}$ ($\lambda^{ija}_{7}$) given in eqs. (\ref{2.4}) and  (\ref{2.2.2}). The asymmetry is generated from interference between tree level decays and one-loop vertex and self-energy diagrams. The one-loop  vertex and self-energy diagrams are shown in Fig. \ref{fig:baryogenesis1}.

The asymmetry is given by
\begin{eqnarray}
\label{asymmetry}
&& \epsilon^{k} = \frac{1}{24 \pi}\frac{\sum_{i,j,l,m,n}{\rm Im}\left[\lambda_{7}^{ijk}\lambda_{7}^{inl\ast} \lambda_{7}^{mjl\ast} \lambda_{7}^{mnk}\right]}{\sum_{i,j} \lambda_{7}^{ijk\ast}\lambda_{7}^{ijk} }  \nonumber\\
&& \times  \left[ {\cal F}_{V}\left(\frac{M^{2}_{N_l}}{M^{2}_{N_k}}\right)+ 3 {\cal F}_{S}\left(\frac{M^{2}_{N_l}}{M^{2}_{N_k}}\right)\right] ,
\end{eqnarray}
where 
\begin{equation}
{\cal F}_{V}=\frac{2 \sqrt{x}}{x-1}, {\cal F}_{S}=\sqrt{x} \; {\rm ln}\left(1+\frac{1}{x}\right).
\end{equation}
${\cal F}_{V}$ corresponds to one-loop function for vertex diagram and  ${\cal F}_{S}$ corresponds to one-loop function for self-energy diagram. The baryon to entropy ratio generated by decays of $N_k$ is given by $n_{B}/s\sim \epsilon \ n_{\gamma}/s \sim  (\epsilon/g_{*})(45/\pi^4)$, where $n_{\gamma}$ is number density of photons per comoving volume and $g_{\ast}$ corresponds to the number of relativistic degrees of freedom. By considering $\lambda^{7}_{ijk} \sim 10^{-3}$ in model 1, one can generate $n_{B}/s\sim 10^{-10}$ for maximal CP violation. Similarly,  one needs $\lambda^{ija}_{7}\sim 10^{-3}$ to satisfy required bound on $n_{B}/s$ in model 2.

In models 3 and 4, $N^{c}_{1,2}$ ($N^{c}$) are Majorana leptons and hence a $B-L$ asymmetry is created via the decay of heavy $N^{c}$ which then gets converted to the baryon asymmetry of the Universe in the presence of the $B+L$ violating anomalous processes before the electroweak phase transition. In these two cases, $N^c_{k (a)}$ decays to the final states $d^{c}_{i} {\tilde h}_{j}, {\tilde d^{c}}_{i} h_{j}$ with $B-L= -1$ and to their conjugate states with $B-L=1$, via the interaction term $\lambda^{ija}_{7}$ ($\lambda^{ijk}_{7}$ ) in Eq. (\ref{2.3.2} (\ref{2.4.2})). The one-loop diagrams that can interfere with the tree level $N_{a} (N_{k})$ decays to provide the required $CP$ violation are again the diagrams given in Fig. \ref{fig:baryogenesis1}. However in these scenarios a $B-L$ asymmetry is created from the decay of Majorana $N^{c}$ in contrast to the $B$ asymmetry created in models 1 and 2.  Again utilizing the general expression for calculating asymmetry parameter  as given in (\ref{asymmetry}), one needs $\lambda^{ija}_{7} (\lambda^{ijk}_{7}) \sim 10^{-3}$ in order to  satisfy ${n_B}/{s}\sim 10^{-10}$ bound in both models 3 and 4.

\begin{figure}[h]
	\includegraphics[width=9cm]{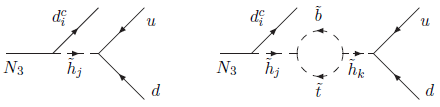}
	\caption{The tree level and one-loop diagrams for $N_{3}$ decay giving rise to baryogenesis in model 5.}
	\label{fig:baryogenesis2}
\end{figure}

For models 5 and 6, we have discussed two possible choices for the second discrete symmetries in section \ref{u1N_models}. In model 5, $N^{c}_{1,2}$ are leptons and $N^{c}_{3}$ is a baryon while in model 6 all the $N^{c}$'s are leptons. For the first choice of second discrete symmetry $Z^{H}_{2}=Z^{L}_{2}$ the form of the superpotential (Eq. \ref{2.5.2} for model 5) clearly shows that one cannot generate the baryon asymmetry of the Universe from high scale leptogenesis via the decay of heavy Majorana $N^{c}$ in these models. However, the term $\l^{ij3}_{7} d^{c}_{i} h_{j} N^{c}_{3}$ can give rise to baryogenesis at $\tev$ scale or below if one consider soft supersymmetry (SUSY) breaking terms  in model 5. The relevant soft SUSY terms in the Lagrangian is given by
\begin{equation}
\label{4.1.5}
{\cal L}\sim  {m^{2}_{{\tilde h}_i}} {{\tilde h}^{\dagger}_i} {{\tilde h}_i} +  {m^{2}_{{\tilde Q}_l}} {{\tilde Q}^{\dagger}_l} {{\tilde Q}_l}+ { A}^{ilm} {{\tilde h}_i}  {{\tilde Q}_l} {{\tilde Q}_m}+...\; \; \; ,
\end{equation}
where $i$ corresponds to the different generations of leptoquarks and  $Q_{l(m)}=(u_{l}, d_{l})$, $l,m=1,2,3$, corresponds to three generations of superpartners of the Standard Model quarks. The Feynman diagrams for the tree level process and the one-loop process interfering with it to provide the $CP$ violation are shown in Fig. \ref{fig:baryogenesis2}.
The  asymmetry parameter in this case is given by \cite{Hambye:2001eu}
\begin{eqnarray}
\epsilon&=& A_{N_3}\sum_{i,j,k}\left[{\rm Im}\left[\lambda^{ij3\ast}_{7}\lambda^{ik3}_{7} {\cal A}^{j33\ast} {\cal A}^{k33}\right]\left(\frac{|\lambda^{j11}_{8}|^{2}}{m^{2}_{\tilde h_j}}-\frac{|\lambda^{k11}_{8}|^{2}}{m^{2}_{\tilde h_k}}\right)\right.\nonumber\\
&& + \left. {\rm Im}\left[\lambda^{ij3\ast}_{7} \lambda^{ik3}_{7} \lambda^{j11}_{8}\lambda^{k11\ast}_{8}\right]\left(\frac{\left|{\cal A}^{j33}\right|^{2}}{m^{2}_{\tilde h_1}}-\frac{\left|{\cal A}^{k33}\right|^{2}}{m^{2}_{\tilde h_1}}\right)\right.\nonumber\\
&& + \left. {\rm Im}\left[{\cal A}^{j33} {\cal A}^{k33\ast} \lambda^{j11}_{8}\lambda^{k11\ast}_{8}\right]\left(\frac{|\lambda^{ij3}_{7}|^{2}}{m^{2}_{\tilde h_j}}-\frac{|\lambda^{ik3}_{7}|^{2}}{m^{2}_{\tilde h_k}}\right) \right]
\end{eqnarray}
where $A_{N_3}=\frac{1}{\Gamma_{N_3}}\frac{1}{(2 \pi)^3}\frac{1}{12}\frac{\pi}{4 \pi^2}\frac{M^{5}_{N_3}}{{m^{2}_{{\tilde h}_j}} m^{2}_{{\tilde h}_k}}$ and $\Gamma_{N_3}$ is the total decay width of $N_3$. Thus, by considering the soft SUSY breaking terms (given in Eq. (\ref{4.1.5})) of $\tev$ scale, one can generate required amount of baryon asymmetry for particular values of Yukawa couplings.

\begin{figure}[h]
	\includegraphics[width=4cm]{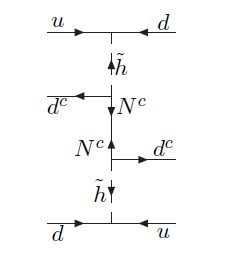}
	\caption{$n$-$\bar{n}$ oscillation induced by effective six-quark interaction.}
	\label{fig:nnbar}
\end{figure}

This can also induce neutron-antinutron ($n$-$\bar{n}$) oscillation violating baryon number by two units ($\D B=2$) \cite{Mohapatra:1980qe}. The effective six-quark interaction inducing $n$-$\bar{n}$ oscillation is shown in Fig. \ref{fig:nnbar}.
In fact, models 1 and 2 can also induce $n$-$\bar{n}$ oscillation in a similar fashion. However in model 6 all the $N^{c}$s are leptons and hence in this model a scheme for baryogenesis similar to above is not possible. 

\begin{figure}[h]
	\includegraphics[width=9cm]{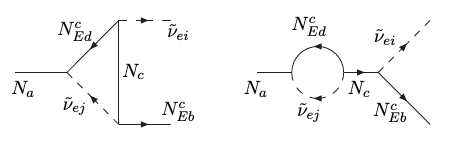}
	\caption{One-loop diagrams for $N_a$ decay which interferes with the tree level decay to provide $CP$ violation.}
	\label{fig:baryogenesis3}
\end{figure}

Now if we choose the second discrete symmetry to be $Z^{H}_{2}=Z^{E}_{2}$ in models 5 and 6 (see Eq. (\ref{2.5.3})) then it is possible to allow high scale leptogenesis via the decay of heavy Majorana $N^{c}$. In these two models $N^{c}_{a (j)}$ decays to the final states $\nu_{e_{i}} {\tilde N}^{c}_{E_{b}}, {\tilde \nu}_{e_{i}} N^{c}_{E_{b}}, e_{i}{\tilde E}^{c}_{b}, {\tilde e}_{i} E^{c}_{b}$ with $B-L= -1$ and to their conjugate states with $B-L=1$, via the interaction term $\lambda^{iab}_{7}$ ($\lambda^{ijb}_{7}$ ) in Eq. (\ref{2.5.4}). Here we take advantage of the fact that $Z^{E}_{2}$ symmetry forbids bilinear term like $LX^{c}$ and consequently $X^{c}$ need not to carry any lepton number, it can simply have the assignment $B=L=0$. The one-loop diagrams for $N_{a} (N_{j})$ decays that can interfere with the tree level decay diagrams to provide the required $CP$ violation are given in Fig. \ref{fig:baryogenesis3}.

For models 7 and 8 the imposition of the $Z^{B}_{2}$ symmetry implies vanishing $\l_{6}$ and $\l_{7}$ for two or more generations of $N^{c}$. Thus in these models no matter what kind of $Z^{H}_{2}$ we choose sufficient $CP$ violation cannot be produced and consequently the possibility of baryogenesis (leptogenesis) from the decay of heavy Majorana $N^{c}$ is ruled out. Thus one needs to resort to some other mechanism to generate the baryon asymmetry of the Universe.
\section{Neutrino masses}
\label{nmass}
In all the variants of $U(1)_{N}$ model that we have considered in Section \ref{u1N_models},  the scalar component of $S_{3}$ acquires a VEV to break the $U(1)_{N}$. The fermionic component of $S_{3}$ pairs up with the gauge fermion  to form a massive Dirac particle. However the fields $S_{1,2}$ still remains massless and can give rise to an interesting neutrino mass matrix structure.

In model 1, the field $N^{c}_{1,2,3}$  are baryons and hence they do not entertain the possibility of canonical seesaw mechanism of generating mass for neutrinos. However, the bilinear terms $\m^{ia}L_{i}X^{c}_{a}$ can give rise to four nonzero masses for $\nu_{e,\m,\t}$ and $S_{1,2}$ as noted in Ref. \cite{Ma:2000jf}. The $9 \times 9$ mass matrix for the neutral fermionic fields of this model $\nu_{e, \mu, \tau}, S_{1,2}, \nu_{E_{1,2}}$ and $N^{c}_{E_{1,2}}$ is given by
\be{\label{4.1}}
\cM^{1}= \bpm 0 & 0 & 0 & \mu^{ia} \\ 0 & 0 & \lambda_{5}^{ab3} v_{2} & \lambda_{5}^{a3b} v_{1} \\  0 & \lambda_{5}^{ba3} v_{2} & 0 & M_{a}\delta_{ab} \\ (\mu^{T})^{ai} &  \lambda_{5}^{b3a} v_{1} & M_{a}\delta_{ab} & 0  \epm,
\ee
where $v_{1}$ and $v_{2}$ are the VEVs acquired by $\tilde{\nu}_{E_{3}}$ and $\tilde{N}^{c}_{E_{3}}$ respectively, and $M_{1,2}$ corresponds to the mass eigenvalues of the neutral fields $X_{1,2}$ and $X^{c}_{1,2}$. We will further assume that the field $\nu_{E_{1,2}}$ pairs up with the charge conjugate states to obtain heavy Dirac mass. Thus in Eq. \ref{4.1} 4 of the 9 fields are very heavy with masses $M_{1}, M_{1}, M_{2}$ and $M_{2}$ to a good approximation. This becomes apparent once we diagonalize $\cM^{1}$ in $M_{a}$ by a rotation about the 3-4 axis to get
\begin{widetext}
\begin{equation}{\label{4.2}}
{\cM^{\prime}}^{1}= \bpm 0 & 0 & \mu^{ia}/\sqrt{2}  & \mu^{ia}/\sqrt{2} \\ 0 & 0 & (\lambda_{5}^{ab3} v_{2}+\lambda_{5}^{a3b} v_{1} )/\sqrt{2} & (-\lambda_{5}^{ab3} v_{2}+\lambda_{5}^{a3b} v_{1} )/\sqrt{2}\\  (\mu^{T})^{ai}/\sqrt{2}& (\lambda_{5}^{ba3} v_{2}+\lambda_{5}^{b3a} v_{1})/\sqrt{2} & M_{a}\delta_{ab}  &0 \\ (\mu^{T})^{ai}/\sqrt{2}  &  (-\lambda_{5}^{ba3} v_{2}+\lambda_{5}^{b3a} v_{1})/\sqrt{2} & 0 & -M_{a}\delta_{ab}  \epm.
\end{equation}
\end{widetext}
Then we readily obtain the $5 \times 5$ reduced mass matrix for the three neutrinos and $S_{1,2}$ given by
\be{\label{4.3}}
\cM^{1}_{\nu}= \bpm 0 & \mu^{ic} \l^{cb3}_{5} v_{2} M_{c}^{-1} \\
\l^{ac3}_{5} \mu^{cj}  v_{2} M_{c}^{-1} & (\l^{ac3}_{5} \l^{c3b}_{5} + \l^{a3c}_{5} \l^{cb3}_{5}) v_{1} v_{2} M_{c}^{-1}  \epm,
\ee
where the repeated dummy indices are summed over. Note that one neutrino remains massless in this model, two of the active neutrinos acquire small masses and the remaining eigenvalues correspond to sterile neutrino states. From Eq. \ref{4.3} it follows that the bilinear terms $\m LX_{c}$ and  the sterile neutrinos are essential for the nonzero active neutrino masses in this model. The fields $N^{c}_{1,2,3}$, which are responsible for creating the baryon asymmetry of the Universe do not enter the neutrino mass matrix anywhere and hence the neutrino masses in this model do not have any direct connection with the baryon asymmetry. To have the active neutrino masses of the order $10^{-4} \ev$ one can choose the sterile neutrino mass of the order $1 \ev$ and the off-diagonal entries in Eq. (\ref{4.3}) to be of the order $10^{-2} \ev$. In this model the oscillations between the three active neutrinos and two sterile neutrinos is natural, and this allows the possibility of accommodating the LSND results \cite{Aguilar:2001ty}. The mixing between $S_{1,2}$ and the heavy neutral leptons $\nu_{E}, N^{c}_{E}$ can give rise to the decays $E_{1,2}\rightarrow W^{-}S_{1,2}$, $E^{c}_{1,2}\rightarrow W^{+}S_{1,2}$, $\n_{{E}_{1,2}}\rightarrow Z S_{1,2}$ and $N^{c}_{1,2}\rightarrow ZS_{1,2}$; which will compete with the decays arising from the Yukawa couplings $E_{1,2}\rightarrow H^{-}S_{1,2}$, $E^{c}_{1,2}\rightarrow H^{+}S_{1,2}$, $\n_{{E}_{1,2}}\rightarrow H^{0} S_{1,2}$ and $N^{c}_{1,2}\rightarrow H^{0}S_{1,2}$, where $H^{+} (H^{0})$ are physical admixture of $\tilde{E}_{3} (\tilde{\n}_{E_{3}})$ and $\tilde{E}^{c}_{3} (\tilde{N}^{c}_{E_{3}})$.

In model 2, $N^{c}_{3}$ is a lepton and hence the term $\l_{6}^{i33} L_{i} N^{c}_{3} X^{c}_{3}$ in the superpotential given in Eq. (\ref{2.2.2}) can give rise to a seesaw mass for one active neutrino, while the other two active neutrinos can acquire masses from Eq. (\ref{4.3}) as before. Thus in this model all three neutrinos can be massive instead of two in model 1. Note that this model can allow the neutrino mass texture where one of the active neutrinos can have mass much larger compared to the other two, which can naturally give atmospheric neutrino oscillations with a $\D m^{2}$ orders of magnitude higher than $\D m^{2}$ for solar neutrino oscillations.

In the case of model 3 all three $N^{c}$ fields are leptons and the $12 \times 12$ mass matrix for the neutral fermions spanning $\nu_{e,\m,\t}$, $S_{1,2}$, $N^{c}_{1,2,3}$, $\nu_{E_{1,2}}$ and $N^{c}_{E_{1,2}}$ is given by
\be{\label{4.3.1}}
\cM^{3}= \bpm 0 & 0 & \l_{6}^{ij3} v_{2} & 0 & \mu^{ia} \\
 0 & 0 & 0 & \lambda_{5}^{ab3} v_{2} & \lambda_{5}^{a3b} v_{1} \\
   \l_{6}^{ji3} v_{2}& 0 & M_{N_{i}} \delta_{ij} & 0 & 0 \\
   0& \lambda_{5}^{ba3} v_{2} & 0 & 0 & M_{a}\delta_{ab} \\
    (\mu^{T})^{ai} &  \lambda_{5}^{b3a} v_{1} & 0 & M_{a}\delta_{ab} & 0  \epm.
\ee
This gives the reduced $5 \times 5$ matrix for three active and two sterile neutrinos as follows
\be{\label{4.3.2}}
\cM^{3}_{\nu}= \bpm \l_{6}^{ik3} \l_{6}^{kj3} v_{2}^{2} M_{N_{k}}^{-1} & \mu^{ic} \l^{cb3}_{5} v_{2} M_{c}^{-1} \\
\l^{ac3}_{5} \mu^{cj}  v_{2} M_{c}^{-1} & (\l^{ac3}_{5} \l^{c3b}_{5} + \l^{a3c}_{5} \l^{cb3}_{5}) v_{1} v_{2} M_{c}^{-1}  \epm.
\ee
This clearly shows that in this model active neutrinos can acquire seesaw masses even in the absence of the bilinear term $\m L X^{c}$ and the sterile neutrinos. As we have discussed in section \ref{leptogenesis}, the out-of-equilibrium decay of $N^{c}$ creates the lepton asymmetry in this model, thus, $M_{N}$ can be constrained from the requirement of successful leptogenesis. However one still has some room left to play with $\l_{5}$, $\m$ and $M_{a}$, which can give rise to interesting neutrino mass textures. In model 4, the fields $N^{c}_{1,2}$ are leptons while $N^{c}_{3}$ is a baryon and hence the $11 \times 11$ mass matrix spanning $\nu_{e,\m,\t}$, $S_{1,2}$, $N^{c}_{1,2}$, $\nu_{E_{1,2}}$ and $N^{c}_{E_{1,2}}$ will reduce to a $5 \times 5$ matrix similar to Eq. (\ref{4.3.2}), except the $(1,1)$ entry which is now given by $\l_{6}^{ic3} \l_{6}^{cj3} v_{2}^{2} M_{N_{c}}^{-1}$. Thus it follows that two of the active neutrinos can acquire masses even without the bilinear term $\m L X^{c}$ and the sterile neutrinos.

For models 5 and 6 we have discussed two possible choices for the second discrete symmetry $Z^{H}_{2}$ in section \ref{u1N_models}. In the former model $N^{c}_{1,2}$ are leptons and $N^{c}_{3}$ is a baryon while in the latter model all $N^{c}_{1,2,3}$ are leptons. In model 5, for the first choice i.e. $Z^{B}_{2}=Z^{L}_{2}$ the $11 \times 11$ mass matrix for the neutral fermions spanning $\nu_{e,\m,\t}$, $S_{1,2}$, $N^{c}_{1,2}$, $\nu_{E_{1,2}}$ is given by
\be{\label{4.5.1}}
\cM^{5}= \bpm 0 & 0 & \l_{6}^{id3} v_{2} & 0 & \mu^{ia} \\
0 & 0 & 0 & \lambda_{5}^{ab3} v_{2} & \lambda_{5}^{a3b} v_{1} \\
\l_{6}^{di3} v_{2}& 0 & M_{N_{d}} \delta_{dg} & 0 & 0 \\
0& \lambda_{5}^{ba3} v_{2} & 0 & 0 & M_{a}\delta_{ab} \\
(\mu^{T})^{ai} &  \lambda_{5}^{b3a} v_{1} & 0 & M_{a}\delta_{ab} & 0  \epm,
\ee
which can be reduced to $5 \times 5$ matrix for 3 active and 2 sterile neutrinos
\be{\label{4.5.2}}
\cM^{3}_{\nu}= \bpm \l_{6}^{ic3} \l_{6}^{cj3} v_{2}^{2} M_{N_{c}}^{-1} & \mu^{ic} \l^{cb3}_{5} v_{2} M_{c}^{-1} \\
\l^{ac3}_{5} (\mu^{T})^{cj}  v_{2} M_{c}^{-1} & (\l^{ac3}_{5} \l^{c3b}_{5} + \l^{a3c}_{5} \l^{cb3}_{5}) v_{1} v_{2} M_{c}^{-1}  \epm,
\ee
which is similar to the form in model 4 and hence similar conclusions follow. Model 6 gives a reduced mass matrix similar to model 3 given in Eq. (\ref{4.3.2}).

For the second choice in model 5, i.e. $Z^{B}_{2}=Z^{E}_{2}$ the $11 \times 11$ mass matrix for the neutral fermions is given by
\be{\label{4.5.3}}
\cM^{5}= \bpm 0 & 0 & 0 & 0 & 0 \\
0 & 0 & 0 & \lambda_{5}^{ab3} v_{2} & \lambda_{5}^{a3b} v_{1} \\
0 & 0 & M_{N_{d}} \delta_{dg} & 0 & 0 \\
0& \lambda_{5}^{ba3} v_{2} & 0 & 0 & M_{a}\delta_{ab} \\
0 &  \lambda_{5}^{b3a} v_{1} & 0 & M_{a}\delta_{ab} & 0  \epm,
\ee
which clearly shows that the active neutrinos are massless in this case while the sterile neutrinos acquire masses  $(\l^{ac3}_{5} \l^{c3b}_{5} + \l^{a3c}_{5} \l^{cb3}_{5}) v_{1} v_{2} M_{c}^{-1}$. The masslessness of the active neutrinos is a consequence of the exotic discrete $Z^{E}_{2}$ symmetry which forbids the mixing among the exotic and nonexotic neutral fermion fields defined in Eq. (\ref{2.5.3}). The situation is similar for $Z^{B}_{2}=Z^{E}_{2}$ in model 6 also.

The analysis of mass matrix for models 7 and 8 are exactly similar to model 1 and 2 respectively with similar conclusions.
\section{Conclusions}
\label{conclusions}
We have studied the variants of effective low-energy $U(1)_{N}$ model motivated by the superstring inspired $E_6$ group in presence of discrete symmetries ensuring proton stability and forbidding tree level flavor changing neutral current processes. Our aim was to explore the eight possible variants  to explain the excess $eejj$ and $e \slashed p_T jj$ events that have been observed by CMS at the LHC and to simultaneously explain the baryon asymmetry of the universe via baryogenesis (leptogenesis). We have also studied the neutrino mass matrices governed by the field assignments and the discrete symmetries in these variants.

We find that all the variants can produce an $eejj$ excess signal via exotic slepton decay, while, the models where $h$ and $h^{c}$ are leptoquarks (models 3, 4, 7 and 8) both $eejj$ and $e\slashed{p}_{T}jj$ signals can be produced simultaneously. For the choice $Z^{H}_{2}=Z^{L}_{2}=(-1)^{L}$ as the second discrete symmetry, two of the variants (model 1 and 2) offers the possibility of direct baryogenesis at high scale via decay of heavy Majorana baryon, while two other (models 3 and 4) can accommodate high-scale leptogenesis. For the above choice of the second discrete symmetry none of the other variants are consistent with high-scale baryogenesis (leptogenesis), however, model 5 allows for the possibility of baryogenesis at $\tev$ scale or below by considering soft supersymmetry breaking terms and this mechanism can induce baryon number violating $n-\bar{n}$ oscillation. On the other hand we have also pointed out a new choice for the second discrete symmetry which has the feature of ensuring proton stability and forbidding tree level FCNC processes, while allowing for the possibility of high scale leptogenesis for models 5 and 6. Studying the neutrino mass matrices for the $U(1)_{N}$ model variants we find that these variants can naturally give three active and two sterile neutrinos and accommodate the LSND results. These neutrinos acquire masses through their mixing with extra neutral fermions and can give rise to interesting neutrino mass textures where the results for the atmospheric and solar neutrino oscillations can be naturally explained.

 
\end{document}